\documentclass[a4paper0,12pt]{article}

\begin{document}

\setlength{\unitlength}{1cm}
\newcommand{\be}{\begin{eqnarray}}
\newcommand{\ee}{\end{eqnarray}}
\newcommand{\bee}{\begin{eqnarray*}}
\newcommand{\eee}{\end{eqnarray*}}
\newcommand{\pmat} {\begin{array}}
\newcommand{\ep} {\end{array}}
\newcommand{\ra}{\rightarrow}
\newcommand{\sse}{\subsection}
\newcommand{\+} {\`}
\newcommand{\es} {\epsilon}
\newcommand{\pa} {\partial}
\newcommand{\ddh} {\dd h}
\newcommand{\al} {\alpha}
\newcommand{\dd} {\dot}
\newcommand{\hb} {\hbar}
\newcommand{\R}{\mbox {\sc R}}
\newcommand{\N}{\mbox {\sc N}}
\newcommand{\Z}{\mbox {\sc Z}}
\newcommand{\C}{\mbox {\sc C}}
\newcommand{\D}{\mbox {\sc D}}
\newcommand{\I}{\mbox {\sc 1}}
\newcommand{\T}{\mbox {\sc T}}
\newcommand{\s}{\mbox {\sc P}}
\newcommand{\p}{\mbox {\sc T}}
\newcommand{\0}{\mbox {\sc 0}}
\newcommand{\Rp}{\mbox {\sc r}}
\newcommand{\Np}{\mbox {\sc n}}
\newcommand{\Zp}{\mbox {\sc z}}
\newcommand{\Cp}{\mbox {\sc c}}
\newcommand{\asy}{{\cal O}

\newcommand{\ind}{\hskip 0.5cm}}
\title{Localization of the states of a $PT$-symmetric double well}
\author{\textbf{Riccardo Giachetti}\\Dipartimento di Fisica e Astronomia, Universit\+a di Firenze,\\ 50019 Sesto Fiorentino.  I.N.F.N. Sezione di Firenze. \\\\ \textbf{Vincenzo Grecchi}\\Dipartimento di Matematica, Universit\+a di Bologna, 40126 Bologna.\\ I.N.F.N. Sezione di Bologna.}
\maketitle
\textbf{Abstract}\\We make a nodal analysis of the processes of level crossings in a  model of quantum mechanics with a \textit{PT}-symmetric double well. We  prove the existence of infinite crossings with their  selection rules. 
At the crossing, before the PT-symmetry breaking and the localization, we have a total P-symmetry breaking of the states.
 \\ 
\section{Introduction}
The interest on simple quantum mechanical models is also given by certain similarities with quantum field theory. In particular, is of great interest
the possibility of summing divergent perturbation series. This problem is related to the existence of singularities of the levels as functions of the perturbation parameter. Such singularities, due to the level crossings, are not so easy to study in a rigorous way. The semiclassical theory provided good qualitative an quantitative results for lower semiclassical parameter up to the crossing value \cite{BW}-\cite{BG}, \cite{BOG}, \cite{A}. The  exact semiclassical method \cite{DDP} has given good qualitative and quantitative results for larger values of the parameter \cite{DP}, \cite{DT}. We believe  that only the nodal analysis, begun in the papers \cite{SHA}, \cite{E}, \cite{EGS}, \cite{GM}, can give  a clear and exhaustive analysis of the level crossings. \\
In recent times the main  interest focused on the $PT$-symmetric Hamiltonians
\cite{BG}, \cite{BB}. In particular it was of some interest to prove the reality and the analyticity of the spectrum of certain oscillators  \cite{SH}, \cite{C}.  Andr\'e Martinez and one of us (V. G.) in the paper  \cite{GM} have proved the Pad\'e summability of the perturbation series  to the perturbative levels $\tilde{E}_n(\beta)$ of  the imaginary cubic oscillator,\be H(\beta)=p^2+x^2+i\sqrt{\beta}x^3,\,\,p^2:=-\frac{d^2}{dx^2},\,\,\beta\neq 0, \,\,|\arg(\beta)|<\pi\,.\label{HB2}\ee 
In this paper \cite{GM}, was also used  the semiclassical method, but the exact results was mostly given by the control of the nodes of the states. Our program is to extend the analysis of the perturbative levels to the other regions of $\beta$ where  the level crossings are expected. \\By changing representation, we study the spectrum of a semiclassical Hamiltonian,  the closed, \textit{PT}-symmetric
    operator, \be\,\,\,H_\hb:=\hb^2p^2+V(x),\,\,V(x):=i(x^3- x),\,\,\, \hb>0, \,\,\,\,\,\,\,\label{RKHKC1}\ee   where the derivative of the potential $V'(x)$ has  two  real zeros:   
     $x_\pm=\pm 1/\sqrt{3}$. In some sense,  (\ref{RKHKC1}) is a  \textit{PT}-symmetric double well Hamiltonian. The large $\hb$ behavior of the levels 
     ${E}_n(\hb)$, corresponding to the perturbative levels $\tilde{E}_n(\beta)$, is studied by the other Hamiltonian,\be K(\al)=p^2+W(\al, x),\,\,\,W(\al,x)=i(x^3+\al x),\,\,\,\,\,\al\geq 0.\label{KH}\ee 
    The level $\hat{E}_m(\al)$, $m\in\N,$ of $K(\al)$ is holomorphic on the sector,
\be\C_\al:=\{\al\in\C; \al\neq 0, |\arg(\al)|< 4\pi/5\},\label{ST}\ee
but can be analytically continued to $-\al_n>-\al>0$ small, up to the first crossing at $\al_n$, where we define directly the level of $H_\hb$, by the relation,
\be E_m(\hb)=\hb^{6/5}\hat{E}_m(\al), \,\,m\in\N,\,\,\,\al=-\hb^{-4/5}\,\,.\label{RHT}\ee
The level $\hat{E}_m(\al)$, real analytic  for $\al>0$ \cite{SH}, \cite{GM}, is  real analytic also for $-\al>0$ small.
  Since we know the absence  of singularities of $E_n(\hb)$ for small  $|\hb|$, or $\hat{E}_n(\al)$ for large $|\al|$ \cite{C}, in certain sectors, we define two other types of levels  for small $\hb>0,$ by the analytic continuation of $\hat{E}_n(\al)$ on the complex plane, along  paths starting from $\al=\hb^{-4/5}$, continuing with $|\al|=\hb^{-4/5},$ and arriving to $\al^\pm:=\exp(\pm i\pi)\hb^{-4/5}$, respectively.
Thus, we define the levels,
\be E_n^\pm(\hb):=\hb^{6/5}\hat{E}_n(\al^\mp), \,\,n\in\N,\,\,\,\al^\pm=\exp(\pm i\pi)\hb^{-4/5},\,\,\,\hb>0.\,\,\label{RHT}\ee
All such levels are extendible as multi-valued functions, to the sector on the $\hb$ complex plane:
 \be\C^0=\{\hb\in \C; \,\hb\neq 0, \,|\arg(\hb)|<\pi/4\}.\label{SIH}\ee 
 We extend all the states, $$\psi_n^\pm (\hb,z) ,\,\,\,n\in\N,\,\,\hb>0\,\, \textrm{small},\,\,\psi_m(\hb,z),\,\,m\in\N,\,\,\,\hb>0\,\,\textrm{large},\,\,z\in\C,$$  for fixed $\hb$, as entire functions on the complex $z$ plane. For large fixed $\hb>0,$  the representation of the state $\psi_m(z)$  is taken  $P_xT$-symmetric, where $P_x\psi_m(x+iy)=\psi_m(-x+iy).$  In this case, the set of the nodes, as the set of the other zeros,  of $\psi_m(z)$ is $P_x$ symmetric. For small fixed $\hb>0,$ we have $\psi_n^+ (z)=P_xT\psi_n^-(z),$ so that the set of nodes $S_n^\pm$ of the state $\psi_n^\pm$ is the $P_x$ transform of $S_n^\mp$. 
For small $\hb>0,$ both the levels $E_n^\pm(\hb),$ $n\in\N,$ $E_n^+(\hb)=\bar{E}_n^-(\hb)$, are non-real up to the first crossing. In particular, the  perturbative levels have the following semiclassical behavior,\be E_n^\pm(\hb)=\mp i\frac{2}{3\sqrt{3}}+ \sqrt{\pm i}\sqrt[4]{3}(2n+1)\,\hb+O(\hb^{2}),\,\,\hb>0,\label{0B}\ee in the limit $\hb\ra 0^+.$ These behaviors correspond to the behaviors of the
perturbative levels of the Hamiltonian (\ref{HB2}). 
In the same limit, all the nodes of the states $\psi_n^\pm(\hb)$  shrink to the centers of the wells, $$x_\pm\in\C^\pm=\{z\in \C; \pm\Re z>0\},$$
respectively.
Now, we prove  that the zeros, and in particular the nodes, of a state $\psi_n^\pm(\hb)$ cannot reach and cross the imaginary axis (Lemma 5). Moreover, the nodes  on $\C^\pm$ respectively, cannot disappear going to infinite (Lemma 4). This means that the number $n$ of the nodes of a state $\psi_n^\pm(\hb)$ in $\C^\pm$
are stable for small $\hb$ up to the first crossing at $h_n=(-\al_n)^{-5/4}>0$. 
At the limit of $\hb\ra h_n^-,$ the energy levels 
$E_n^\pm(\hb)$ have the limit 
$ E_n^c>0$ and the set of the zeros of $\psi_n^\pm(\hb)$ becomes $P_x$-symmetric.  The critical state $\psi_n^c$, has a $P_x$-symmetric set of 2n nodes in $\C^{+-}:=\C^+\bigcup\C^-$ (Lemma 7). The discontinuity of the number of nodes at $h_n$, $n>0,$ is due to the \textit{PT}-symmetry breaking of the states.\\
Now we look for a pair  of positive analytic levels $E_j(\hb),E_k(\hb)$, $j,k\in\N$, $j\neq k$, with limits $E_j(h_n^+)=E_k(h_n^+)=E_n^c$ and corresponding states with the limit value $\psi_j(h_n^+)=\psi_k(h_n^+)=\psi_n^c$.   
 This is possible if both the corresponding $P_x$-symmetric sets of nodes  $S_j(\hb),S_k(\hb)$ contain  $2n$ nodes  stable in  $\C^{+-}$.  
 But what can be said about the imaginary nodes?  In order to distinguish a possible node in the imaginary axis from the other zeros, we consider the limit $\hb\ra+\infty$ corresponding to the limit $\al\ra 0^-.$ We prove (Lemma 6) that the  nodes in this limit are confined in the lower half plane. Since the imaginary turning point,
 for  a level $E>0$, is $I_0=i\tilde{y}$, $\tilde{y}>0,$  we define as an imaginary node a zero in
 $\Sigma=i(-\infty,\tilde{y})$. We also prove that  the number of imaginary nodes can be zero or one. Since there are in any case $2n$ non imaginary nodes,  the two independent states  are necessarily the states  $(\psi_{2n}(\hb),\psi_{2n+1}(\hb))$, with levels $(E_{2n}(\hb)<E_{2n+1}(\hb))$ for $\hb>h_n$.  Actually, exists $h_n^p>h_n$ such that the state $\psi_{2n+1}(\hb)$  has an imaginary node for $\hb>h^p_n$.  In Lemma 9 we give the crossing selection rules (\ref{CR}):
     $$ E_n^\pm(h_n^-)=E_m(h_n^+):=E_n^c>0,$$\be\,\,\psi_n^\pm(h_n^-)=\psi_m(h_n^+):=\psi_n^c ,\,\,\forall m\in\N,\,\, [m/2]=n.\ee 
The critical state $\psi_n^c$ is orthogonal to its $P$-transform, or, in other words,  is totally $P$-asymmetric (Lemma 10).
The crossing corresponds to a square root singularity
of this pair of analytic functions, positive analytic for $\hb>h_n.$ 
For the value of $h^p_n$ and $E_{2n+1}(h_n^p):=E^p_{n}$, we have only numerical results.
 Our numerical computations (Table 1) show that 
$h^p_n\ra 0$, and $E_{2n+1}(h_n^p):=E^p_{n}\ra E^p>0$ as $n\ra\infty$ with $E_n^p-E^p=O((h_n^p)^2)$, where  $E^p\sim 0,352268..$  is the unique energy such that the imaginary turning point \cite{DT} is on the short Stokes line \cite{E}, \cite{EGS}.   The semiclassical state corresponding to the energy $E^p$ is considered a bilocalized state, the transition state between the delocalized and the localized state.  \\
 The unicity of the crossing for each pair $E_n^\pm$ is taken as a conjecture (Conjecture 1) assumed in order to simplify the notations and the discussion.
\\
We give the structure of the Riemann 
sheet of the levels $E_m(\hb)$ from large $\hb>0$ to all the real axis, with the values at the borders of the cut $(0,h_n]$ (Theorem 1).\\In Sec. 2 we prove the positivity of the spectrum for large $\hb$ and the reality of the states on the imaginary axis; in Sec. 3 we consider the appearance of an imaginary node for the odd states;
In Sec. 4, we follow the process of crossing;   in Sec. 5 we prove that for small $\hb>0$ the imaginary axis is free of zeros and the nodes are bounded; in Sec. 6 we prove a confinement of the nodes for large $\hb>0$; in Sec. 7 we  prove the quantization rules,   the continuity and the boundedness of the levels; in Sec. 8 we prove the total P-symmetry breaking at the crossing; in Sec. 9  we give  the local structure of the Riemann sheets of the positive levels with the cuts directed toward $ 0$.
 \begin{table}[h]
\begin{small}
\begin{center}
\begin{tabular}{c|c|c|c}
n & {$h_n^p$}  & {$E_n^p$}\\
\hline
8 & 0.043835&  0.3519 {$\pm$}0.0010 \\
9 & 0.030683 &  0.3514 {$\pm$} 0.0011 \\
10 & 0.023605 &  0.3518{$\pm$} 0.0013 \\
11& 0.013060 &  0.3522 {$\pm$} 0.0002 \\
\hline
\end{tabular}
\end{center}
\end{small}
\caption{ The values of  $h_n^p$, and $E_n^p$ with the errors, for different values of $n=8-11$. }
\end{table}\\
\section{Positivity of the levels  and reality of the states on the imaginary axis for large $\hb>0$} The  level  $\hat{E}_m(\al)$,  $m\in\N$ of $K(\al)$ is  analytic in a neighborhood  of the origin $U\subset \C$ \cite{GM}, \cite{S}. Since it is real analytic   for $\al<0$ it is real analytic also in  $U\bigcup\R$ \cite{SH}.  The positivity comes from the
       positivity of the kinetic energy, 
       $$\Re \hat{E}_m(\al)=\Re <\widehat{\psi}_m(\al),K(\al)\widehat{\psi}_m(\al)>=<\widehat{\psi}_m(\al),p^2\widehat{\psi}_m(\al)>\,>\,0,$$ where $\psi_m(\al)$ is the corresponding normalized state. Also the level $E_m(\hb)$ is real analytic and positive
     for $\hb>0$ large enough.
Thus, we have proved:\\
 \textbf{Lemma 1}\\ \textit{The level $E_m(\hb)$,  $m\in\N,$ is real analytic and positive for  $\hb>0$ large enough.}\\\\
 We now extend the analysis of the analytic states on the complex plane.\\Let us consider $y\in\R$ and the  translation  $ f(x)\ra f(x+iy)$, so that the \textit{PT}-symmetric Hamiltonian    becomes the other {isospectral} \textit{PT}-symmetric Hamiltonian on the $\mathcal{H}_y$ representation:
 \be H_\hb(y):=h^2p^2+i(x^3-(3y^2+1)x)-(3yx^2-y^3-y)\sim H_\hb.\label{RKHK1}\ee
 The eigenfunction $\psi_{n,y}(x):=\psi_n(x+iy)$ on the $\mathcal{H}_y$ representation, with {real eigenvalue $E_n$, can be  taken $PT$-symmetric}:\be PT\psi_{n,y}(x):=\overline{\psi}_{n,y}(-x),\label{PX}\ee and in particular $$\,\,\,\psi_{n,y}(0)
 =\overline{\psi}_{n,y}(0)=\psi(iy).$$   
 Thus, we have proved the following,
 \\\textbf{Lemma 2}\\
 \textit{For large $\hb>0,$ the level $E_m(\hb)$, $m\in \N,$ is positive and, for a choice of the gauge, the state $\psi_m(\hb)$, extended to the complex plane as an entire function, is $P_xT$-symmetric: \be (P_xT\psi_m)(x+iy):=\overline{\psi}_m(-x+iy)={\psi}_m(x+iy),\,\,\,\forall x,y\in\R,\label{PX1}\ee and, in particular, the state is real on the imaginary axis,  \be \Im\psi_m(iy)=0,\,\,\forall y\in\R.\label{RSI1}\ee The set of all the zeros  of the state is $P_x$-symmetric.}
\section{The nodal analysis of the process of crossing }
Let  $E_m(\hb),$ for $\hb>0$ large enough,  be a positive level  of the Hamiltonian (\ref{RKHKC1}) with a corresponding state $\psi_m(\hb)$. Now, by the complex dilation $z\ra iz$, we consider  the  Hamiltonian on the imaginary axis:
 \be   H^r_\hb=-\hb^2\frac{d^2}{dy^2}+\tilde{V}(y)\sim-H_\hb,\,\,\,
 \tilde{V}(y):=-y^3-y,\label{MHB}\ee  {well defined} by the $L^2$ condition on the $x$-axis, here playing the role of the imaginary axis. The Hamiltonian $H^r_\hb$  has the same  spectrum as $-H_\hb,$  so that  $-E:=-E_m(\hb)<0$ is one of its eigenvalues. The corresponding  state  $\phi_m(y):=\psi_m(iy)$  can be taken  real. Actually, since we can have the $P_xT$-symmetry of the entire state $\psi_m(z),$ we can have the reality of $\psi(iy)$ for real $y$: $\psi_m(x+iy)=\bar{\psi}_m(-x+iy)$,
 $\psi_m(iy)=\bar{\psi}_m(iy)$.
  We consider together the two states $\psi_m(z)$,    $[m/2]=n\in\N$,  for a fixed $\hb\geq h_n$ (\ref{CR}). Both the states have  $n$ nodes on both the half-planes $\C^\pm$ and are distinguished by the number of imaginary nodes  for $\hb>0$ large.
All the process of crossing for $\hb\geq h_n$ can be studied by the behaviors of the states $\psi_m(z)$, with energy $E=E_m$, $[m/2]=n$ on the imaginary semi-axis, \be\Sigma(E)=\{z=iy; -\infty<y<\tilde{y}(E)\},\label{CBOIA}\ee where the imaginary turning point is $I_0=i\tilde{y}(E)$. This means that we consider the two states $\phi_m(y)$, of (\ref{MHB}), with energies $-E=-E_m$, $[m/2]=n$ for $y\leq {\tilde{y}}(E)$.\\  For $\hb>0$ large, we have two possible behaviors of the state $\phi(y)$ of (\ref{MHB}) with level $-E$. Let us recall that if, for $y$ in a bounded interval of the semi-axis $-\infty<y<\tilde{y}(E)$, a state $\phi(y)$ is positive, it is  convex;
 if it is negative, it is  concave. On the other side,  for $y>\tilde{y}(E)$, where an eigenfunction $\phi(y)$ is positive  it is also concave,
and where it is negative it  is also convex.\\  Since we can consider $\phi(y)$
positive decreasing for $y<<\tilde{y}(E),$ there are only two cases: \\a) the existence of one  node  on $\Sigma(E)$,  \\b) the  absence of nodes on $\Sigma(E)$. \\Let us remark that  $\tilde{y}(E)>0$  so that a possible  node on the imaginary axis should be in $\Sigma(E)$ for large $\hb$.
Thus, we have the result:\\
\textbf{Lemma 3}\\\textit{The state $\psi_m(\hb)$, $m\in\N$, with corresponding positive level $E=E_m(\hb)$,  have at most one zero in $\Sigma (E)$.  This zero, considered a node, exists  for $\hb>h_n$ large enough if  $m=2n+1$, but  don't exists if $m=2n$. }
\section{ Non real levels:   imaginary axis   free of zeros and bounded nodes }
 Let us consider the general case with a level $E\in\C,$ and the corresponding  state $\psi(z),$ with $\psi(iy)=\phi(y),$ $z,y\in \C$.
 \ We transform the Hamiltonian by imaginary translations: $$ H^r_\hb(x)=-\hb^2\frac{d^2}{dy^2}+\tilde{V}(y-ix),\,$$
 $$\,\tilde{V}(y-ix)=-(y-ix)^3-(y-ix)=-y^3+3x^2y-y+i(x(3y^2+1)-x^3)=$$$$=\Re \tilde{V}(y-ix)+i\Im \tilde{V}(y-ix),$$
   where $\Im \tilde{V}(y-ix)=(x(3y^2+1)-x^3)$ with level $-E$, for a fixed $x\neq 0,$ and we consider a state, $$\phi_x(y):=\phi(y-ix),\,\,\,\,\,n\in\N,$$
  with the well known asymptotic behaviors,$$|\phi_x(y)|^2\sim C|y|^{-3/2}\, \,\,\textrm{for}\,\,\,\, y\ra+\infty,$$\be|\phi_x(y)|^2\sim C|y|^{-3/2}\exp(-2y^{5/2}/\hb) \,\,\textrm{for}\,\, y\ra -\infty\,\,.\label{AA}\ee
 We apply the Loeffel-Martin method \cite{LM} generalized to the case of diverging integrals:
 $$\hb^2\Im \,[\overline{\phi}_x(y)\,\pa_y\phi_x(y)]=$$
 \be=\hb^2\Im \,[\overline{\phi}_x(\hat{y})\pa_y\phi_x(\hat{y})]+
 \int_{\hat{y}}^y(x(3s^2+1)-x^3+\Im E)
 |\phi_x(s)|^2ds\ra+\infty,\label{AN}\ee 
 where $x(3s^2+1)-x^3=\Im \tilde{V}(s-ix)$,  as $y\ra+\infty$ for fixed $\tilde{y},x\in\R,$ $x\neq 0.$ We know that   the zeros, for $|z|$ large, have the asymptotic direction $\arg z\ra \pi/2$ \cite{GM}. By (\ref{AN}) we prove a stronger condition on the asymptotics of the zeros:
\\ \textbf{Lemma 4} \\\textit{Let  $E$ be a level with state $\psi(z)$ of the Hamiltonian $H_\hb$ for a fixed $\hb>0$. Consider a generic  zero  $Z_j=X_j+iY_j$ of $\psi(z)$. Exists $M>0$, such that   $\pm X_j>0$  if $\mp\Im E>0,$ $Y_j>M$. In case of real level, $\Im E=0$, the large zeros are purely imaginary. }\\\textbf{proof}\\
The integral in (\ref{AN}) don't diverge for $y\ra+\infty$ only if $x$ depends on $y$ such that, \be x(y)\ra 0, \,\,\pm x(y)\geq \frac{|\Im E|}{3y^2+1},\,\,\,\mp\Im E>0,\,\,\,\textrm{as}\,\,\,y\ra+\infty.\label{AADZ}\ee 
Actually, condition (\ref{AADZ}) is necessary for having a change of sign on the integrand in (\ref{AN}). Otherwise, the integral in (\ref{AN}) diverge.
   We have the state $\phi_E(y):=\psi_E(iy)$ with corresponding  level $E$ of the Hamiltonian $H_\hb$.
 We consider the Loeffel-Martin formula    in order to generalize to our problem the expression of the imaginary part of a shape resonance:
 \be\hb^2\,\Im \,(\overline{\phi}(y)\pa_y\phi(y))
 =\Im E\,\int_{-\infty}^{y}
 |\phi(s)|^2ds\neq 0,\,\,\forall y\in \R,\label{AN1}\ee 
 where the integral in (\ref{AN1}) exists bounded for the semiclassical behavior. Thus, we state the result:
 \\ \textbf{Lemma 5} \textit{Let us consider the level $E=E^\pm_n(\hb)$, for $\hb<h_n$, so that   $\Im E\neq 0.$ The corresponding state  $\psi(z)=\psi_n^\pm(\hb,z)$ 
 is different from $0$ for all  $z=iy,\,y\in\R.$}.
\section{Confinement  of  the nodes for large $\hb>0$}
  We prove a confinement of the nodes for the states in the case of a degenerate double well.  Let us consider  the level $\hat{E}_m(0)$, $m\in\N,$ of $K(\al)$ at $\al=0$, corresponding to the level $E_m(\hb)$ of $H_\hb$ at the limit of $\hb=+\infty$,  because of  the relation (\ref{RHT}), $E_n(\hb)=\hb^{6/5}\hat{E}_n(-\hb^{-4/5})$. It is relevant that the scaling used for this relation (\ref{RHT}) is a regular one  with a positive scale  $\lambda=\hb^{2/5}$ (even if infinite) respecting the angles on the complex plane.  It is known that the level $\hat{E}_m(\al)$ is positive for $\al\geq 0$ \cite{SH}. We prove now a confinement of the zeros which allows us to distinguish the nodes from the other zeros.
\\
We consider the operator $K(0)$ (\ref{KH})  translated by $x\ra x+iy,$
 $$ K_y(0)=p^2+i(x+iy)^3=
 p^2+i(x^2-3y^2)x+y^3-3yx^2:=p^2+V_y(x)\,\,\, .$$
 We apply the Loeffel-Martin method \cite{LM} to a level $E=\hat{E}_m(0)$, with $E>0$:
 $$-\Im \,[\overline{\psi}(x+iy)\pa_x\psi(x+iy)]=\int_{x}^\infty\Im V_y(s)\,|\psi(s+iy)|^2ds=\int_{x}^\infty(s^2-3y^2 )
 s|\psi(s+iy)|^2ds=$$$$=-\int^{x}_{-\infty}(s^2-3y^2 )
 s|\psi(s+iy)|^2ds\neq 0,\,\,$$ for $\pm x\geq \sqrt{3}\,|y|$, $y\in\R$.
 In this case  we have a rigorous confinement, extendible to all $\al>0$, of the region of   the nodes,
 $$\C_\sigma=\{z=x+iy;y<0,|x|<-\sqrt{3}\,y\}\subset \C_-=\{z=x+iy;y<0\}.$$ Since the same confinement extends to all $\al>0,$ we have that the $m$ zeros of the state $\widehat{\psi}_m(\al)$ on $\C_-$ are stable in the limit $\al\ra+\infty,$ i. e. are nodes
 by definition. Previous computations of the nodes \cite{BBS}   suggest that  the present confinement may be  sharp.\\
 Thus, we state a result:\\
 \textbf{Lemma 6}\\\textit{All the nodes of  the state $\psi_m(\hb,z)$, $ m\in\N,$ for  $\hb>0$ large enough, are in $\C_-.$}
\section{The  quantization rules}
Suppose the existence of a continuation of each level $E^\pm:=E_n^\pm(\hb)$ from $\hb<h_n$ to $\hb\geq h_n.$ For the moment, we keep  the same names $E_n^\pm(\hb)$ for the continuations of the levels even if  such names are no more specific. We have two kinds of quantization rules for a fixed $\hb>0$ small, giving the eigenvalues $E_n^\pm(\hb)$, respectively. Exist  two regular circuits $$\gamma^\pm,$$ such that,
 $P_x\gamma^+=\gamma^-$, $$\gamma^\pm=\partial D^\pm ,$$
where $D^\pm$ is a regular region large enough, with $$D^\pm\subset\C^\pm:=\{x+iy, \pm x>0, y\in\R\};$$
and,
\be
\frac{1}{2i\pi}\oint_{\gamma^\pm}\frac{\psi'(\hb,E^\pm,z)}{\psi(\hb,E^\pm,z)}dz=n, \label{CC}\ee
 where $\pm\Im E^\pm\geq 0.$  In particular, for small $\hb>0,$
we have the semiclassical quantization condition, \be
\frac{1}{2i\pi}\oint_{\gamma^\pm}p_0(E^\pm,z)dz=\hb(n+\frac{1}{2})+O(\hb^2). \label{CC1}\ee This quantization conditions are still valid for all $\hb>0,$ but, for large $\hb \geq h_n$, are both  satisfied by both the new states $\psi_m(\hb)$, $m\in\N,$ $[m/2]=n$ and the critical state $\psi_n^c$.
We now prove that the state $\psi_n^c$ has the set of $2n$ nodes in $\C^+\bigcup\C^-$. 
At the limit of $\hb\ra h_n^-,$ the energy levels 
$E_n^\pm(\hb)$ have the limit 
$ E_n^c>0$ and the set of the zeros of $\psi_n^\pm(\hb)$ becomes $P_x$-symmetric. Let a node $N_j\in \C^+$ of $\psi_n^+$ to have a limit $N_j^c$  as $\hb\ra h_n^- $ . Because of the  
$P_x$ symmetry of the set of all the zeros at the limit $\hb\ra h_n^-$,  does  exist a  zero $Z_k$ such that $ Z_k\ra P_xN_j^c$ as 
$\hb\ra h^-_n$. It is possible to disprove the possibility that $\Re N_j^c=0.$ Actually, in this case $N_j^c$ would be a double zero of 
the state $\psi_n^c:=\psi_n^\pm(h_n^-)$, but we know that the zeros are simple.
Thus, the sets $S_n^\pm(\hb)$ of the $n$ nodes of the states $\psi_n^\pm(\hb)$ have  limits $S_n^\pm (h_n^-)\in\C^\pm$, respectively, as $\hb\ra h_n^-$. ).  Both the states $\psi_m(\hb)$, $m\in\N$, $[m/2]=n$ for $\hb>h_n$, have $2n$ nodes in $\C^{+-}$. Actually,  these noses are stable: cannot become imaginary for the symmetry and the simplicity of the spectrum and cannot diverge along the imaginary axis. We have proved the:\\ \textbf{Lemma 7}\\\textit{The critical state $\psi_n^c$, has a $P_x$-symmetric set of 2n nodes in $\C^{+-}:=\C^+\bigcup \C^-$, 
\be S_n^c=S_n^\pm (h_n^-)\bigcup P_x S_n^\pm (h_n^-)=S_n^+ (h_n^-)\bigcup S_n^- (h_n^-).\ee 
 Both the states $\psi_m(\hb)$, $m\in\N$, $[m/2]=n$ for $\hb>h_n$, have $2n$ nodes in $\C^{+-}.$}\\\\
  In order to select a single state it is sufficient to use the inequality $E_{2n+1}(\hb)>E_{2n+1}(\hb)$. One more node of $\psi_{2n+1}(\hb)$ lies  on the imaginary axis. This is clear for large $\hb>0$, where all the nodes have a negative imaginary part. In this case it is possible to give a unique quantization rule,
\be
\frac{1}{2i\pi}\oint_{\Gamma}\frac{\psi'(\hb,E,z)}{\psi(\hb,E,z)}dz=m, \label{CC2}\ee
where $m=2n$ or $2n+1,$ and $\Gamma=\partial \Omega$ for $\Omega\subset \C_-,$ $\C_-=\{z\in\C; \Im z<0\}.$
    \\ For $\hb>0$ large, it is convenient to use the scaling of the operator $K(\al)$ in order to have  energy and nodes uniformly bounded.
    \\ \textbf{Lemma 8}\\\textit{For each $n\in\N$, does exist $h_n>0$ and a crossing:}
     $$ E_n^\pm(h_n^-)=E_m(h_n^+):=E_n^c>0,$$\be\,\,\psi_n^\pm(h_n^-)=\psi_m(h_n^+):=\psi_n^c ,\,\,\forall m\in\N,\,\, [m/2]=n.\label{CR}\ee  \textbf{Proof}\\The existence of this crossing is necessary because of the positivity of the analytic functions $E_m(\hb)$ for large $\hb>0,$ and the non reality of the analytic functions $E_n^\pm(\hb)$ for small $\hb>0.$ The relation between the integer $n$ and the integers $m$ is due to the doubling of the nodes at $h_n^-$ (Lemma 4) because of the $P_x$ symmetry of the set of nodes. This nodes are in $\C^{+-}$ and are stable for $\hb\geq h_n.$ The differentiation of the number of nodes of the two states in necessary for large $\hb>0$. Actually, for $\hb>h_n^o>h_n$  
     $\psi_{2n+1}(\hb)$ has imaginary node $N_0=iy$ with $y<\tilde{y}$, where the imaginary turning point is $I_0=i\tilde{y}$. \\\textbf{Conjecture 1}\\\textit{We assume that  the crossing between the pair $E_n^\pm$, $\forall n\in\N$, is unique}.\\This conjecture is justified by numerical, semiclassical and exact semiclassical results \cite{A} \cite{ZJ}, \cite{DT}. This conjecture allow us to simplify the notations.\\ 
  Now we prove that each level is bounded for bounded parameter $\hb>0.$\\
 \\ {\textbf{Lemma 9}}\\\textit{Let $E(\hb)$,  be any level of the pair $E_n^\pm(\hb),$ for $\hb<h_n$,  or  any level of the pair $E_m(\hb)$, $m\in\N$, $[m/2]=n$ for $\hb>h_n$. Does not exists a $h^c\geq 0,$ such that $E(\hb)$ diverge as $\hb\ra h^c$.}\\
 \textbf{Proof}\\We prove by absurd, and we consider the case $E=E_m(\hb)$ for a fixed $m\in\N,$ and $\hb$ near $h^c>0$. The extension to the general case is simple. We consider the operator $$\frac{H_\hb-E(\hb)}{|E|(\hb)}\sim {\dot h}^2p^2+ix^3-i\al x-\eta,$$ by a scaling $x\ra\lambda x,$ $\lambda=|E|^{1/3}$, where $\dot h=\hb |E^{-5/3}|,$ $\al=|E|^{-2/3}$, $\eta=E/|E|$, $|\eta|=1$.  For small $\dot h>0,$ by  simply putting $\al=0,$
we have the semiclassical quantization condition, \be
\frac{1}{2i\pi}\oint_{\Gamma}\sqrt{\eta -iz^3}dz=\dot h(m+\frac{1}{2})+O((\dot h)^2). \label{CC2}\ee which can be valid only if
$\eta\ra 0$ as $\dot h\ra0.$
\\\section{At the crossing  the state is orthogonal to its P-transform}
 We have a crossing of $E_n^\pm(\hb)$ at $\hb=h_n$, when $\Im E_n^\pm(\hb)=0.$ For $0<\hb<h_n,$ the two clamped points of $\psi_n^\pm$ are $(I_\mp,I_0)$ respectively.
 At the crossing, we have $P_x$ symmetry of the turning points, so that $I_-=\bar{I_+},$ $I_0=-\bar{I_0}$.\\
 Let   $H:=H_\hb$,  $H_\hb^*=\bar H:=H_{\bar{\hb}}$, with two eigenvalues $E_j=\bar E_j$, and eigenvectors $\psi_j$ $j=1,2$. We have $$H\psi_1=E_1\psi_1,\,\,\,\, \bar H\bar\psi_2=E_2\bar\psi_2,$$ so that \be<\bar\psi_2,H\psi_1>=E_1<\bar\psi_2,\psi_1>,\label{PE}\ee
$$<H\psi_1,\bar\psi_2>=<\psi_1,\bar H\bar\psi_2>=E_2<\psi_1,\bar\psi_2>,$$  and, by complex conjugation,\be<\bar \psi_2,H\psi_1>=E_2<\bar\psi_2,\psi_1>.\label{SE}\ee By subtraction of the two equations (\ref{PE}) (\ref{SE}), we get,$$0=(E_2-E_1)<\psi_1,\bar\psi_2>.$$ Let now to vary the semiclassical parameter $\hb$, so that:
 $$ 0=(E_2(\hb)-E_1(\hb))<\psi_1(\hb),\bar\psi_2(\hb)>,$$ for $\hb>0.$ If $E_1(\hb)\neq E_2(\hb)$ for $\hb> h_n,$  and $E_1(h^+_n)=E_2(h^+_n)=E,$ $\psi_1(h^+_n)=\psi_2(h^+_n)=\psi,$ we have \be 0=<\psi,\bar\psi>=<\psi,P\psi>=\int_{\R }\psi^2(x)dx.\label{CA}\ee
\\Thus, we have proved the following:
 \\ \\\textbf{Lemma 10}\\\textit{
 The PT-symmetric state at the crossing point, $${\psi}^c_n=\psi_{2n+1}(h^+_{n})=\psi_{2n}(h^+_{n})=PT{\psi}^c_n,$$
    is completely P-asymmetric, i.e. is orthogonal to its P-transform: $$\int_{\R} {\psi}^c_n\,(x)^2dx=<{\psi}^c_n,P{\psi}^c_n>=0.$$}
\section{The Riemann surface  near the real axis}
Let us consider the sector on the $\hb$ complex plane (\ref{SIH}):
$$\C^0=\{\hb\in \C;\,\hb\neq 0\,, \arg(\hb)<\pi/4\},$$ and the Riemann sheet $\C^0_m$ of the level $E_m(\hb)$, $n=[m/2]$, defined  in $\C^0,$
 with a square root singularity at $h_n$
 and a cut, $\gamma_{n,n}=(0,h_n]$. Thus, we prove the following:\\\\
\textbf{Theorem 1}\\\textit{ 
The levels $(E_{2n+1}(\hb),E_{2n}(\hb)),$ $n\in\N,$ are analytic functions defined on the Riemann sheets $(\C^0_{2n},\,\,\C^0_{2n+1})$, respectively, both with  only the cut  $\gamma_{n,n}=(0,h_n]$ on the real axis. The  
positive analytic functions $(E_{2n+1}(\hb),E_{2n}(\hb)),$   with $E_{2n+1}(\hb)>E_{2n}(\hb)$ on $(h_n,+\infty)$ have the following values at the borders of the cut: \be E_{2n}(\hb\pm i0^+)={E}^\pm_{n}(\hb),\,\,E_{2n+1}(\hb\pm i0^+)={E}^\mp_{n}(\hb),\,\,\forall\,\, 0<\hb<h_n.\label{SON}\ee}
\textbf{Proof}\\ Since both the functions $(E_{2n+}(\hb),E_{2n}(\hb))$ have a square root singularity at $\hb_n,$ and $$E_{2n+1}(\hb_n+\es)-E_{2n}(\hb_n+\es)=O(\sqrt{\es})>0,$$ for $\es>0$ small, $$\pm\Im(E_{2n+1}(\hb_n+\exp(\pm i\pi)\es)-E_{2n}(\hb_n+\exp(\pm i\pi)\es))<0,$$ and $\mp\Im E_n^\pm (h)>0,$ for $h<h_n,$ necessarily we have, $$E_{2n+1}(\hb_n+\exp(\pm i\pi)\es):={E}^\mp_{n}(\hb_n-\es),$$ $$E_{2n}(\hb_n+\exp(\pm i\pi)\es):={E}^\pm_{n}(\hb_n-\es).$$
 \\
 \textbf{Remark }\\The Riemann sheet $\C^0_0$ of the fundamental level has only one cut $\gamma_{0,0}=[0,h_0]$ on $\R$ \cite{DT}, and the discontinuity on the cut is defined by  the rule, \be E_{0}(\hb\pm i0^+)={E}^\pm_{0}(\hb),\,\,\forall\,\,\hb,\,\, 0<\hb<h_0.\label{S0N}\ee
 We recall, for instance, that $E_0^+(\hb):=E_0(\hb+i0)$, is defined as the limit from above for small $\hb>0.$ This definition extends directly to all $\hb>0$ in absence of complex singularities. Formula (\ref{S0N})  means that it is possible the absence of other singularities involving the function $E_0(\hb)$. Thus, using the principle of maximal analyticity, we assume that in $\C_0^0$ there is only the cut $\gamma_{0,0}$.\\
\\\textbf{Aknowlodgments}\\It is a pleasure to thanks Professor Andr\'e Martinez for long and useful discussions at the beginning of this research.


\begin{thebibliography}{99}
\bibitem {SH} Shin, K C,
\textit{On the reality of eigenvalues for a class of PT-Symmetric oscillators},
Commun. Math. Phys. \textbf{104},
229 (3), 543-564 (2002).
\bibitem {GM} Grecchi V, Martinez A, \textit{The Spectrum of the Cubic Oscillator} Commun. Math. Phys. \textbf{319}  479-500 (2013).\\see also:\\
  Grecchi V, Maioli M,  Martinez A: \textit{Pad\'e summability of the cubic oscillator}, J. Phys. A: Math. Theor. \textbf{42}  425208 (17 pp) (2009);\\
    V. Grecchi, M. Maioli, A. Martinez, \textit{The top resonances of the cubic oscillator}, J. Phys. A: Math. Theor.  \textbf{43} n.47 (2010).
\bibitem {BW} Bender C M, Wu T T,  \textit{Anharmonic oscllator}, {Phys. Rev.} \textbf{184} 1231-60 (1969).
    \bibitem {SI} Simon B Ann. Phys. \textbf{58}, 76 (1970).
    \bibitem {HS} Harrel E M II, Simon B Duke Math. j B \textbf{47},47 (1980).
    \bibitem {BG} Benassi L, Grecchi V \textit{Resonances in the Stark effect and strongly asymptotic approxiamnts} J. Phys. B: At. Mol. Phys. \textbf{13},  911 (1980).
\bibitem {SHA} Shanley P E, \textit{Spectral properties of the scaled quartic anharmonic oscillator} Ann. Phys. (N.Y.) \textbf{186},  292-324.  Shanley P E, \textit{Nodal properties of the quartic anharmonic oscillator} Ann. Phys. (N.Y.) \textbf{186}, 325-354 (1988).
    \bibitem {E}
    A. Eremenko, A. Gabrielov: \textit{Analytic continuation of eigenvalues of a quartic oscillator}, Comm. Math. Physics \textbf{287},  431-457 (2009).
\bibitem {EGS} A. Eremenko, A. Gabrielov, B. Shapiro: \textit{Zeros of eigenfunctions of some anharmonic oscillators}. Ann. Inst. Fourier, \textbf{58},  603-624 (2008);
\textit{High energy eigenfunctions of one-dimensional Schrodinger operators with polynomial potentials}. Comput. Methods and Function Theory, \textbf{8}  513-529 (2008).\\
\bibitem {A} Alvarez G, \textit{Bender-Wu branch points in the cubic oscillator} \textit{J. Phys. A: Math. Gen.} \textbf{27} 4589-4598 (1995).
     \bibitem {DP} Delabaere E,  Pham F: \textit{Unfolding the quartic oscillator} Ann. Phys. NY \textbf{261} 180-218 (1997).
         \bibitem {DT} Delabaere E, Trinh D T, \textit{Spectral analysis of the complex cubic oscillator} J. Phys. A: Math. Gen. \textbf{33}  8771-8796 (2000).
     \bibitem {DDP} Delabaere E, Dillinger H, Pham F: \textit{Exact semiclassical expansions for one-dimensional quantum oscillators} J.  Math. Phys. \textbf{38} (12)   6126-6184 (1997).
        \bibitem {ZJ} Zinn-Justin J, Jentschura U D: \textit{Imaginary cubic perturbation: numerical and analytic study}, J.  Phys. A: Math. Phys. \textbf{75}  425301 (29 pp) (2010).
    \bibitem {BOG} Bouslaev V, Grecchi V: \textit{Equivalence of unstable anharmonic oscillators
and double wells},
J. Phys. A Math. Gen. \textbf{26},  5541-5549 (1993).
\bibitem {BB} Bender C M and Boettcher S : \textit{Real Spectra in Non-Hermitian Hamiltonian
Having PT Symmetry},
 Phys. Rev. Lett. \textbf{80}, 5243 (1998).
 \bibitem {BBS} Bender C M, Boettcher S and Savage V M: \textit{Conjecture on interlacing of
zeros in complex
Sturm-Liouville problems }, J. Math. Phys. \textbf{41}, 6381-6387 (1999).
\bibitem {C} Caliceti E, J. Phys. \textbf{A 33}  3753 (2000).
\bibitem {LM} Loeffel J,Martin A, Simon B and Wightman A,  Phys. Lett. B \textbf{30} 656 (1969).
\bibitem {S} Sibuya Y: \textit{Global theory   of a second order linear ordinary differential
equation with a polynomial coefficient,} Chap. 7, Math. Studies 18, North Holland, (1975).

\end{thebibliography}
\end {document}